\documentclass[aps,groupedaddress,prd,eqsecnum]{revtex4}

\usepackage{amsmath,amssymb}
\usepackage{color}

\begin{document}

\title{Black hole entropy for the general area spectrum}

\author{Tomo Tanaka}
\affiliation{Department of Physics, Waseda University, Okubo 3-4-1, Tokyo 
169-8555, Japan}
\email{tomo@gravity.phys.waseda.ac.jp}
\author{Takashi Tamaki}
\affiliation{Department of Physics, Waseda University, Okubo 3-4-1, Tokyo 
169-8555, Japan}
\affiliation{Department of Physics, Rikkyo University, Toshima, Tokyo 171-8501, Japan}
\email{tamaki@gravity.phys.waseda.ac.jp}


\begin{abstract}

We consider the possibility that the horizon area is expressed by 
the general area spectrum in loop quantum gravity and calculate 
the black hole entropy by counting the degrees of freedom in spin-network states 
related to its area. Although the general area spectrum has a complex expression, 
we succeeded in obtaining the result that the black hole entropy is proportional to 
its area as in previous works where the simplified area formula has been used. 

The meaning of this result is important since we can reconfirm the idea that 
the black hole entropy is related to the degrees of freedom in spin-network states. 
We also obtain new values for the Barbero-Immirzi parameter 
($\gamma =0.5802\cdots \ \mathrm{or} \ 0.7847\cdots$) which are larger than 
that of previous works. 

\end{abstract}

\date{\today}

\maketitle


\section{Introduction}

Statistical mechanics in a self-gravitating system is quite different 
from that without gravity. For example, particles in the box have maximal 
entropy when they spread out uniformly in the box if gravity is not taken 
into account. On the other hand, if particles are self-gravitating, we can 
suppose that clusters appear as an entropically favorable state. Then, if 
the pressure of particles can be neglected, it is likely that a black hole 
appears as a maximal entropy state. Thus, black hole entropy 
would be the key for understanding statistics in a self-gravitating system. 

One of the most mysterious things about black holes is their entropy $S$ which 
is {\it not} proportional to its volume {\it but} to its horizon area $A$. 
This was first pointed out related to the first law of black hole 
thermodynamics~\cite{Bekenstein}. The famous relation $S=A/4$ 
has been established by the discovery of the Hawking radiation~\cite{Hawking}. 
Recently, its statistical origin has been discussed in 
the candidate theories of quantum gravity, such as string theory~\cite{Strominger}, 
or loop quantum gravity (LQG)~\cite{Rovelli-entropy}, etc. It has been discussed 
that LQG can describe its statistical origin independent of 
black hole species because of its background independent formulation~\cite{ABCK}. 
For this reason, we concentrate on LQG here. 

Quantum states in LQG are described by spin-network \cite{Smolin}, 
and basic ingredients of the spin-network are edges, 
which are lines labeled by spin $j$($j=0,1/2,1,3/2,\cdots$) reflecting 
the SU(2) nature of the gauge group, and vertices which are intersections between edges. 
For three edges having spin $j_1,j_2, \ {\rm and}\  j_3$ that merge at an 
arbitrary vertex, we have following conditions. 
\begin{eqnarray}
 && j_1 + j_2 +j_3 \in \mathbb{N}, \label{integer} \\
 && j_i \leqq j_j + j_k, \ (i,j,k \ \rm{different \ from \ each \ other.}). 
 \label{spincondi}
\end{eqnarray}
These conditions guarantee the gauge invariance of the spin-network. 

Using this formalism, general expressions for the spectrum of the area and the 
volume operators can be derived \cite{Rovelli,Ash1}. 
For example, the area spectrum $A_j$ is 
\begin{eqnarray}\label{area}
 A_j=4\pi \gamma \sum_i \sqrt{2j_{i}^{u}(j_{i}^{u}+1)+2j_{i}^{d}(j_{i}^{d}+1)-j_{i}^{t}(j_{i}^{t}+1)}\ , 
\label{generalized}
\end{eqnarray}
where $\gamma$ is the Barbero-Immirzi parameter related to an ambiguity 
in the choice of canonically conjugate variables \cite{Immirzi}. 
The sum is added up all intersections between a surface and edges. 
Here, the indices $u$, $d$, and $t$ mean edges 
upper side, down side, and tangential to the surface, respectively (We can determine 
which side is upper or down side arbitrarily). 

In \cite{Rovelli-entropy}, it was proposed that black hole entropy is obtained 
by counting the number of degrees of freedom about $j$ when we fix the horizon area 
where a simplified area formula is used. 
This simplified area formula is obtained by assuming that 
there are no tangential edges on black hole horizon, that is $j^t_i=0$. 
We obtain $j^u_i=j^d_i:=j_i$ by using the condition (\ref{spincondi}).
Then, we consider the degrees of freedom about $j$ satisfying 
\begin{eqnarray}\label{simpleareacondi}
 A_j = 8\pi\gamma \sum_i \sqrt{j_i(j_i+1)} = A .\label{simple}
\end{eqnarray}
The standard procedure is to impose the Bekenstein-Hawking entropy-area 
law $S=A/4$ for macroscopic black holes in order to fix the value of $\gamma$. 
Ashtekar et al. in \cite{ABCK} extended this idea using the isolated horizon 
framework (ABCK framework)~\cite{isolated}. Error in counting in this 
original work has been corrected in \cite{Domagala,Meissner}. 
Similar works appear related to how to count the number of freedom in 
\cite{Alekseev,Khriplovich,Mitra,Tamaki2,Sahlmann}. 

However, should we restrict to the simplified area spectrum (\ref{simple}) ?
Thiemann in \cite{Thiemann} used the boundary condition that there is 
no other side of the horizon, i.e., $j_i^d=0$. Then, by using (\ref{spincondi}), 
we obtain $j_i^u=j_i^t:=j_i$ which gives 
\begin{eqnarray}
A_j=4\pi \gamma \sum_i \sqrt{j_{i}(j_{i}+1)}\ . 
\label{simplified2}
\end{eqnarray}
Based on this proposal, the number counting has been performed in \cite{Tamaki3} 
which gives $\gamma =0.323\cdots $. 

Another interesting possibility is to use (\ref{generalized}) which we discuss in 
this paper. In \cite{Rovelli-entropy}, it has been argued that the since the horizon fluctuates, 
we can neglect the possibility that the vertex is on the horizon. This means that 
$j^t_i=0$ resulting the formula (\ref{simpleareacondi}). This is intuitively plausible. 
However, we should distinguish the vertex from the point at the spacetime. Actually, 
if we review the conditions (\ref{integer}) and (\ref{spincondi}) 
that should be satisfied at the vertex, we notice that 
the vertex increases the number of degrees of freedom compared with that without vertex. 
Thus, it is not evident whether 
we can neglect the vertex on the horizon or not. In this sense, it is important 
to consider (\ref{generalized}) as the horizon and determine the number of states. 

Moreover, we can discuss (\ref{generalized}) for the horizon motivated by the hypothesis 
that a black hole is a maximal entropy state in a self-gravitating system. 
If we discuss the microscopic process of the self-gravitating system, 
it is appropriate to imagine evolution of spin-network states. 
Then, since the black hole would appear as a final stage, we should consider 
its corresponding in spin-network states. 
If we agree that the origin of the black hole entropy is related to degrees of freedom 
in $j$ (or $m=-j,-j+1,\cdots,j$ considered in \cite{ABCK}), it is evident that 
(\ref{generalized}) can gain larger number of states than (\ref{simple}) or (\ref{simplified2}) for 
the fixed area. Therefore, if we consider evolution of spin-network states, 
the horizon might appear as a coarse graining of vertices with approximately spherical distribution. 
See, also \cite{Ansari} which also discuss using (\ref{generalized}) as 
expressing the horizon area. 

Of course, it is speculative and the typical objection to the idea is that 
since the black hole evaporates, it is not the maximal entropy state. 
However, the black hole we consider is the limit $A\to\infty$ where the evaporation 
process can be negligible. The second objection is that if we require the entropy-area law $S=A/4$, 
the black hole entropy does not depend on what types of area formula we use, so 
it is not relevant to the above hypothesis. This is a delicate question to be answered 
carefully. From the view point that the Barbero-Immirzi parameter is determined {\it a priori}, 
the formula $S=A/4$ only provides us the method to {\it know} the value of $\gamma$. 
If this is the case, using (\ref{generalized}) would enhance the entropy. 
Therefore, we concentrate on evaluating the number of states using 
(\ref{generalized}) by adopting this view point. 
To answer whether this view point is true or not, we need independent 
discussion to {\it know} the value of $\gamma$ through, e.g., cosmology \cite{LQC} or 
quasinormal modes of black holes \cite{Schiappa,Dreyer,Hod}. 

Our strategy is as follows. Based on the observation that the value of $\gamma$ in 
\cite{ABCK} is qualitatively same as that inferred in \cite{Rovelli-entropy} which 
counts the degrees of freedom of $j$ without imposing the horizon conditions 
for the case (\ref{simple}), we restrict counting the corresponding $j$ freedom for 
(\ref{generalized}) as a first step. We can perform it by carefully reanalyzing 
the case (\ref{simple}). This paper is organized as follows. In section II, 
we review how to count the degrees of 
freedom for (\ref{simple}). In section III, we extend its method for the case (\ref{generalized}). 
In section IV, we mention concluding remarks.

\section{Revisiting the simplified area spectrum }

Here, we show how to count the number of states about $j$ in the simplified area spectrum 
based on \cite{Domagala,Meissner} where counting $m$ freedom have been considered. 
See, also \cite{Diaz-Polo} for another efficient method to count the number of states. 
We consider the following number of states $N(A)$ : 
\begin{eqnarray}
 N(A) := \left\{ (j_1,\cdots ,j_n) | 0 \neq j_i \in \frac{\mathbb{N}}{2} ,\ \ 
	 \sum_i \sqrt{j_i(j_i+1)} =  \frac{A}{8\pi\gamma} \right\}. 
\end{eqnarray}
We derive a recursion relation to obtain the value of $N(A)$. 
When we consider $(j_1, \cdots ,j_n)\in N(A-a_{1/2})$ 
we obtain $(j_1, \cdots ,j_n,\frac{1}{2})\in N(A)$, 
where $a_{1/2}$ is the minimum area where only one $j=1/2$ edge contributes 
to the area eigenvalue (\ref{simple}), i.e., 
$a_{1/2}=8\pi\gamma\sqrt{\frac{1}{2}(\frac{1}{2}+1)}=4\pi\gamma\sqrt{3}$. 
Likewise, for any eigenvalue $a_{j_{x}}$($0 < a_{j_{x}} \leqq A$) of the area operator, we have 
\begin{eqnarray}\label{counting2}
(j_1, \cdots ,j_n)\in N(A-a_{j_{x}}) 
\Rightarrow (j_1, \cdots ,j_n,j_x)\in N(A). 
\end{eqnarray}
For $j_{x}\neq j_{x'}$, we have 
\begin{eqnarray}\label{counting1}
(j_1, \cdots ,j_n,j_x)\neq (j_1, \cdots ,j_n,j_{x'}). 
\end{eqnarray}
Then, important point is that 
if we consider all $0 < a_{j_{x}} \leqq A$ and $(j_1, \cdots ,j_n)\in N(A-a_{j_{x}})$, 
$(j_1, \cdots ,j_n,j_x)$ form the entire set $N(A)$. 

From (\ref{counting2}) and (\ref{counting1}), we obtain 
\begin{eqnarray}
 N(A) = \sum_{j} N(A-8\pi\gamma\sqrt{j(j+1)}).   
 \label{pre-recursion}
\end{eqnarray}
To generalize this formula for (\ref{generalized}) is our main task. 

\section{Consideration of the general area spectrum}
In the case for (\ref{simple}), it has been shown that isolated horizon conditions do not affect 
the number of states in the limit $A\to\infty$. Based on this observation, 
we consider only degrees of freedom about its area (\ref{generalized}) as a first step. 
Then in this case, we also denote number of states as $N(A)$ which is defined as 
\begin{eqnarray}
\hspace{-10mm}N(A) := &&
\left\{(j^u_1,j^d_1,j^t_1,\cdots,j^u_n,j^d_n,j^t_n)|0 \neq j^u_i, j^d_i 
 \in \frac{\mathbb{N}}{2},\ \ 0 \neq j^t_i \in \mathbb{N},
 \ \ j^u_i, j^d_i, j^t_i\ {\rm should\ satisfy\ (\ref{integer})\ and\ (\ref{spincondi}).}  
 \right.  \nonumber  \\\quad 
\hspace{-10mm}&& \left. \sum_i\sqrt{2j_i^u(j_i^u+1)+2j_i^d(j_i^d+1)-j_i^t(j_i^t+1)} =
 \frac{A}{4\pi\gamma} \right\}\ .
\end{eqnarray}
We adopt the condition $j^{t}\in \mathbb{N}$ motivated by the ABCK framework 
where the ``classical horizon'' is described by $U(1)$ connection. 
This is, of course, not verified in the present situation and should be 
reconsidered in future. 

Then, we perform counting as follows. If we use the condition $j^{t}\in \mathbb{N}$, 
we have $j^u+j^d:=n\in \mathbb{N}$ by (\ref{integer}). 
If we fix $n$, we can classify the possible $j^{u}$, $j^{d}$, $j^{t}$ as follows, 
which is one of the most important parts in this paper. 
First, we have $(j^u ,j^d)=(\frac{n}{2}\pm\frac{s}{2},\frac{n}{2}\mp\frac{s}{2})$ 
(double-sign corresponds) for $0\leqq s \leqq n$, $s\in \mathbb{N}$ to satisfy (\ref{spincondi}). 
Then, for each $s$, possible value of $j^t$ is $j^t =s,s+1,\cdots ,n$ to 
satisfy (\ref{spincondi}). This relation is summarized schematically as follows: 
\begin{eqnarray*}
(j^u,j^d)\ \ \ \ \ \ \ \  =(n,0)\ \ \ \ \ \ \  &\rightarrow& j^t=n \\
\vdots\ \ \ \ \ \ \ \ \ \  && \ \ \ \ \ \ \ \ \ \ \ \vdots \\
=(\frac{n}{2}+\frac{s}{2},\frac{n}{2}-\frac{s}{2}) &\rightarrow& \ \ \  =s,s+1,\cdots ,n \\
\vdots\ \ \ \ \ \ \ \ \ \  && \ \ \ \ \ \ \ \ \ \ \ \vdots \\
=(\frac{n}{2}+\frac{1}{2},\frac{n}{2}-\frac{1}{2}) &\rightarrow& \ \ \  =1,2,\cdots ,n \\
=(\frac{n}{2},\frac{n}{2})\ \ \ \ \ \  &\rightarrow& \ \ \ =0,1,\cdots ,n \\
=(\frac{n}{2}-\frac{1}{2},\frac{n}{2}+\frac{1}{2}) &\rightarrow& \ \ \  =1,2,\cdots ,n \\
\vdots\ \ \ \ \ \ \ \ \ \  && \ \ \ \ \ \ \ \ \ \ \ \vdots \\
=(\frac{n}{2}-\frac{s}{2},\frac{n}{2}+\frac{s}{2}) &\rightarrow& \ \ \  =s,s+1,\cdots ,n \\
\vdots\ \ \ \ \ \ \ \ \ \  && \ \ \ \ \ \ \ \ \ \ \ \vdots \\
=(0,n) \ \ \ \ \ \ \   &\rightarrow& \ \ \  =n .
\end{eqnarray*}
Corresponding to (\ref{counting2}), 
for any eigenvalue $x:=4\pi\gamma\sqrt{2j^u_x (j^u_x+1)+
2j^d_x (j^d_x +1)-j^t_x (j^t_x +1)}$ ($0 <x\leqq A$) of 
the area operator, we have 
\begin{eqnarray}\label{counting4}
({\bf j}_1,\cdots ,{\bf j}_n)\in N(A-x) 
\Rightarrow ({\bf j}_1,\cdots ,{\bf j}_n,{\bf j}_x)\in N(A), 
\end{eqnarray}
where we used the abbreviation as ${\bf j}_{i}=(j^u_i,j^d_i,j^t_i)$. 
Corresponding to (\ref{counting1}), we have 
\begin{eqnarray}\label{counting3}
({\bf j}_1,\cdots ,{\bf j}_n,{\bf j}_x)\neq 
({\bf j}_1,\cdots ,{\bf j}_n,{\bf j}_{x'}), 
\end{eqnarray}
if ${\bf j}_x\neq {\bf j}_{x'}$. 

Therefore, as for the case in (\ref{simple}), if we consider all $0 < x \leqq A$ and 
$({\bf j}_1,\cdots ,{\bf j}_n)\in N(A-x)$, $({\bf j}_1,\cdots ,{\bf j}_n,{\bf j}_x)$ 
form the entire set $N(A)$. 

Then, if we use the notation $j^u=\frac{n}{2}+\frac{s}{2},j^d=\frac{n}{2}-\frac{s}{2},j^t=t$, 
we have $x(n,s,t)=4\pi\gamma\sqrt{n^2+2n+s^2-t(t+1)}$ and 
\begin{eqnarray}\label{recursion-general}
 N(A)=\sum^{\infty}_{n=1}\left[\sum^n_{s=1}\sum^n_{t=s}2N(A-x(n,s,t))+
 \sum^n_{t=0}N(A-x(n,s=0,t))\right] ,
\end{eqnarray}
where the factor $2$ in front of $N(A-x(n,s,t))$ for $s\neq 0$ 
corresponds to the fact that same $x(n,s,t)$ appears twice for the exchange of 
$j^u$ and $j^d$. For $A\to\infty$, by assuming the relation: 
\begin{eqnarray}\label{number}
 N(A) = C e^{\frac{A\gamma_M}{4\gamma}} , 
\end{eqnarray}
where $C$ is a constant and substituting to the recursion 
relation (\ref{recursion-general}), we obtain the beautiful formula as a 
generalization of the case (\ref{simple}) as, 
\begin{eqnarray}
 1=\sum^{\infty}_{n=1}\left[\sum^n_{s=1}\sum^n_{t=s}2\exp(-\gamma_Mx(n,s,t)/4\gamma)+
 \sum^n_{t=0}\exp(-\gamma_Mx(n,s=0,t)/4\gamma)\right]. 
\label{beautiful}
\end{eqnarray}
If we require $S=A/4$, we have $\gamma = \gamma_M = 0.5802 \cdots$. 
This means that even if we use (\ref{generalized}) as the horizon spectrum, 
we can reproduce the entropy formula $S=A/4$ by adjusting the Barbero-Immirzi parameter. 
This is nontrivial and is our main result in this paper. 

Let us turn back to our assumptions. Although we obtained $\gamma$ satisfying $S=A/4$ 
for the case (\ref{generalized}), there may be a criticism that the result is 
underestimated by only counting $j$ freedom. To answer it, we consider the following counting. 
When the simplified area formula was used, there is an proposal that 
we should count not only $j$ but also $m=-j,-j+1,\cdots, j$ freedom based on the 
ABCK framework~\cite{Mitra}. Although it is nontrivial 
whether this framework can be extended to the general area formula, 
let us count also the $m$ freedom for each $j^u$ to maximize the 
estimate. Counting only $m$ related to $j^u$ is reasonable 
from the point of view of the entanglement entropy~\cite{Nielsen,Bombelli} or the 
holography principle~\cite{'tHooft}. See, also \cite{Livine,Donnelly} for applying 
the entanglement entropy in LQG context. 

If we notice that there are $(n+s+1)$ and $(n-s+1)$ freedoms for $m$ (total 
$2(n+1)$) corresponding to $(j^{u},j^{d})=(\frac{n}{2}+\frac{s}{2},\frac{n}{2}-\frac{s}{2})$ 
and $(\frac{n}{2}-\frac{s}{2},\frac{n}{2}+\frac{s}{2})$, respectively, 
the factor $2$ in the first term of the right-hand side of (\ref{beautiful}) 
is replaced by $2(n+1)$ in this case. For $s=0$, the factor $1$ in the second term 
is replaced by $(n+1)$. Then, we obtain 
\begin{eqnarray}
 1=\sum^{\infty}_{n=1}\left[\sum^n_{s=1}\sum^n_{t=s}2(n+1)\exp(-\gamma_Mx(n,s,t)/4\gamma)+
 \sum^n_{t=0}(n+1)\exp(-\gamma_Mx(n,s=0,t)/4\gamma)\right],
\end{eqnarray}
which gives $\gamma=\gamma_M=0.7847\cdots$. Thus, we confirm that the black hole 
entropy is proportional to the area again. 
Naively speaking, we expect that there is no qualitative deviation from 
these two values of $\gamma$ even if we take into account the 
ABCK framework for (\ref{generalized}) appropriately. 

\section{Conclusion and Discussion}

In this paper, we obtained the black hole entropy by considering the general area formula. 
It is surprising that we succeeded in obtaining the black hole entropy 
proportional to the horizon area even in this case. 
Moreover, our result shows that using the general area formula highly increases 
the number of degrees of freedom, which suggests that it would be easier to be 
realized compared with the simplified area formula. 
However, we should discuss this feature more carefully. 
There are many possibilities 
examining the area spectrum. For example, we have not yet established the black hole 
thermodynamics in LQG which is one of the most important topics to be investigated. 
There is an idea that black hole 
evaporation process should also be described by using the general area 
formula~\cite{Ansari}. Therefore, whether we can establish the generalized second law 
of black hole thermodynamics might be one of the criteria in judging which area 
formula is appropriate. For this purpose, it is desirable to extend the ABCK framework 
for the general area formula since the exact counting is required. 
Though we do not take care of the topology of the horizon, discussing the 
difference caused by the topology is important as considered for 
the simplified area formula~\cite{Kloster}. 
The covariant entropy bound~\cite{Bousso} is also 
important which has been discussed in the LQG context recently~\cite{Wilson}. 

Of course, as we mentioned in the introduction, 
we should check the value of the Barbero-Immirzi parameter in 
several independent discussions. 
Therefore, we should also take care of cosmology \cite{LQC} and 
quasinormal modes of black holes \cite{Schiappa,Dreyer,Hod} 
in determining the Barbero-Immirzi parameter. Confirming LQG in many 
independent methods would be the holy grail of the theory.

\section*{Acknowledgements}

We would like to thank Tomohiro Harada and Kei-ichi Maeda 
for useful comments and continuous encouragement.


\end{document}